\documentstyle[aps,preprint]{revtex}

\begin{document}
\draft
\title{Damping of low-energy excitations of a trapped Bose condensate at finite
temperatures}
\author{P.O. Fedichev$^{1,2}$, G.V. Shlyapnikov$^{1,2}$, and J.T.M. Walraven$^1$}
\address{$^1$ FOM Institute for Atomic and Molecular Physics,\\
Kruislaan 407, 1098 SJ Amsterdam, The Netherlands}
\address{$^2$ Russian Research Center Kurchatov Institute, Kurchatov\\
Square, 123182 Moscow, Russia}
\date{\today}
\maketitle

\begin{abstract}
We present the theory of damping of low-energy excitations of a trapped Bose
condensate at finite temperatures, where the damping is provided by the
interaction of these excitations with the thermal excitations. We emphasize
the key role of stochastization in the behavior of the thermal excitations
for damping in non-spherical traps. The damping rates of the lowest
excitations, following from our theory, are in fair agreement with the data
of recent JILA and MIT experiments. The damping of quasiclassical
excitations is determined by the condensate boundary region, and the result
for the damping rate is drastically different from that in a spatially
homogeneous gas. 
\end{abstract}

\vspace{4mm} \narrowtext

\tightenlines

After the discovery of Bose-Einstein condensation (BEC) \cite
{Cor95,Hul95,Ket95}, one of the major directions in the physics of
ultra-cold gases has been the investigation of collective many-body effects.
Especially interesting is the behavior of low-energy collective excitations
of a trapped condensate. The JILA \cite{jila,Jin96} and MIT \cite{mit,Ket97}
experimental studies of the excitations related to shape oscillations of the
condensate show that these excitations are damped and provide us with
interesting results on the temperature dependence of the damping rates and
frequency shifts.

In this letter we develop the theory of damping of excitations of a trapped
condensate in the Thomas-Fermi regime at finite temperatures, where the
presence of a thermal component is important. We confine ourselves to the
damping of low-energy excitations, i.e., the excitations with energies $%
E_{\nu}\ll\mu$, where $\mu$ is the chemical potential, and consider
temperatures $T\gg\hbar\omega$ ($\omega$ is the characteristic trap
frequency) ranging almost up to the BEC transition temperature $T_c$. Thus
far, theoretical and numerical investigations of elementary excitations of
trapped Bose-condensed gases predominantly remained on the mean field level 
\cite{Stringari,Burnett,Singh,Jav,Perez,You,Castin,Kagan97,Ohberg,Fantoni}.
The investigation of damping phenomena requires analysis beyond the ordinary
mean field approach\cite{fn1}. It should be emphasized that the damping of
low-energy excitations in a trapped Bose-condensed gas differs fundamentally
from the damping of Bogolyubov excitations in an infinitely large spatially
homogeneous gas. In the latter case, characterized by a continuum of
excitations, any given excitation can decay into two excitations of lower
energy and momentum. This damping mechanism, first discussed by Beliaev for $%
T=0$ \cite{Bel} and employed by Popov \cite{Popov} at finite temperatures,
proves to be dominant at $T\ll\mu$. In a trapped Bose-condensed gas the
character of the discrete structure of the spectrum of low-energy
excitations makes Beliaev damping impossible under conservation of energy.
Therefore, irrespective of the relation between $T$ and $\mu$, the damping
of excitations with energies 
\begin{equation}  \label{ET}
E_{\nu}\ll\mu,T
\end{equation}
has to be provided by their interaction with the thermal excitations. The
damping mechanism involves processes in which the low-energy excitation ($\nu
$) and the thermal excitation ($\gamma$) are annihilated (created) and
another thermal excitation ($\gamma^{\prime}$) is created (annihilated): 
\begin{equation}  \label{process}
\nu + \gamma \leftrightarrow \gamma^{\prime}.
\end{equation}
We will discuss the case where the thermal excitations $\gamma$, $%
\gamma^{\prime}$ are in the collisionless regime. Under the condition (\ref
{ET}) the energies $E_{\gamma}$ of these excitations are much larger than
the energies $E_{\nu}$ of the low-energy excitations. Therefore, the damping
mechanism governed by the processes (\ref{process}) can be treated as Landau
damping. For spatially homogeneous gases this mechanism was first discussed
by Szepfalusy and Kondor \cite{SK,f22}.

It is worth noting that inside the condensate spatial region, at $T\!\alt%
\!\mu$ the density of occupied states of thermal excitations peaks at the
energies $E_{\gamma}\!\sim\!T$, whereas for $T\!\gg\!\mu$ this happens at $%
E_{\gamma}\!\sim\!\mu$. As just the excitations with $E_{\gamma}\!\sim\!\mu$
give the main contribution to the damping rate, the collective character of
the thermal excitations remains important even at $T\gg\mu$ (cf. \cite{SK}).

In a trapped Bose-condensed gas the damping of low-energy excitations is
determined by the behavior of the wavefunctions and by the distribution of
the level spacings of thermal excitations with energies $E_{\gamma}\alt\mu$,
which depends on the trap symmetry. We emphasize that stochastization in the
behavior of these thermal excitations plays a key role for damping in
non-spherical traps. For quasiclassical ($E_{\nu}\gg\hbar\omega$) low-energy
excitations  the main contribution to the damping rate $\Gamma_{\nu}$ comes
from the boundary region of the condensate, which makes $\Gamma_{\nu}$
completely different from that in a spatially homogeneous gas. The damping
of the lowest excitations ($E_{\nu}\sim\hbar\omega$) is determined by the
behaviour of the excitations in the entire condensate region. For this case
the damping rates following from our theory are in fair agreement with the
data of the JILA \cite{Jin96} and MIT \cite{Ket97} experiments.

Elementary excitations of a Bose condensate trapped in an external potential 
$V({\bf r})$ are commonly defined within the Bogolyubov-De Gennes approach
(see \cite{Gennes}) on the basis of the grand canonical Hamiltonian 
\begin{equation}  \label{H}
\!\hat H\!\!=\!\!\!\int\!\!\!d{\bf r}\hat\Psi^{\dagger}\!({\bf r})\!\!\left[
\!-\frac{\hbar^{2}}{2m}\Delta\!+\!V({\bf r})\!+\!\! \frac{\tilde U}{2}\!\hat%
\Psi^{\dagger}\!({\bf r}) \hat\Psi({\bf r})\!-\!\mu\!\right]\!\hat\Psi({\bf r%
})\!\!\!
\end{equation}
assuming a point interaction between atoms, with $\tilde U%
\!=\!4\pi\hbar^2\!a\!/\!m$, $m$ the atom mass, and $a$ the (positive)
scattering length. The field operator of atoms $\hat\Psi$ is represented as
the sum of the above-condensate part $\hat\Psi^{\prime}$ and the condensate
wavefunction $\Psi_0\!=\!\langle\hat\Psi\rangle$. Omitting the terms
proportional to $\hat\Psi^{\prime 3}$ and $\hat\Psi^{\prime 4}$ and using
the generalized Bogolyubov transformation $\hat\Psi^{\prime}({\bf r}%
)=\sum_{\nu}\hat b_{\nu} u_{\nu}({\bf r}) -\hat b_{\nu}^{\dagger}
v_{\nu}^{*}({\bf r})$, where $\hat b_{\nu}$, $\hat b_{\nu}^{\dagger}$ are
annihilation and creation operators of elementary excitations, the
Hamiltonian (\ref{H}) is reduced to the diagonal form $\hat H\!=\!\hat H%
_{0}\!+\!\sum_{\nu}E_{\nu}\hat b^{\dagger}_{\nu}\hat b_{\nu}$, if the
functions $u_{\nu}$, $v_\nu$ satisfy the equations 
\begin{equation}  \label{uv}
\!\!\!\left(\!\!{\frac{{-\hbar^{2}\!\Delta}}{{2 m}}}\!+\!V\!({\bf r}%
)\!-\!\mu\!\!\right)\!\!\!\left[\!\matrix{u_{\nu}\cr v_{\nu}}\!\!\right]\!\!
+\!\tilde U\! |\Psi_{0}|^{2}\!\!\left(\!\! 2\!\left[\!\matrix{u_{\nu}\cr
v_{\nu}}\!\!\right]\!\!-\!\! \left[\!\matrix{v_{\nu}\cr u_{\nu}}%
\!\!\right]\! \right)\!\!=\!\!E_{\nu}\!\left[\!\matrix{u_{\nu}\cr -v_{\nu}}%
\!\!\right]\!\!\!
\end{equation}
The condensate wavefunction $\Psi_0$ is determined by the well-known
Gross-Pitaevskii equation. In the Thomas-Fermi regime, where $%
\!\mu\!\approx\! n_{0m}\tilde U$ ($n_{0m}$ is the maximum condensate
density) greatly exceeds $\hbar\omega$, one has \cite{Silvera,Huse}: $\Psi_0(%
{\bf r})\!=\![(\mu\!-\!V({\bf r}))/\tilde U]^{1/2}$ for $\mu\!\geq\!V({\bf r}%
)$ and zero otherwise. The low-energy excitations ($E_{\nu}\!\ll\!\mu$) are
localized in the condensate spatial region, and in a harmonic potential $V(%
{\bf r})$ the functions $f_{\nu}^{\pm}\!=\!u_{\nu}\!\pm\!v_{\nu}$ can be
written as \cite{Ohberg}: 
\begin{equation}  \label{fpm}
f_{\nu}^{\pm}=\left[\frac{2n_0({\bf r})\tilde U}{E_{\nu}} \right]^{\pm
1/2}\!\!\!\left(\prod_il_i\right)^{\!-1/2}\!\!\!\!\! W_{\nu}(r_i/l_i),
\end{equation}
where $n_0({\bf r})=|\Psi_0({\bf r})|^2$ is the condensate density, $%
l_i=(2\mu/m\omega_i^2)^{1/2}$ the characteristic size of the condensate in
the $i$-th direction, and $\omega_i$ the $i$-th trap frequency.

Interaction between the excitations, caused by the terms proportional to $%
\hat\Psi^{\prime 3}$ and $\hat\Psi^{\prime 4}$ in Eq.(\ref{H}), leads to
damping. Below we will assume the inequality 
\begin{equation}  \label{small}
(T/n_{0m}\tilde U)(n_{0m}a^3)^{1/3}\ll 1
\end{equation}
which is fulfilled up to $T\approx 0.9T_c$ for the conditions of the JILA 
\cite{Jin96} and MIT \cite{Ket97} experiments. Just Eq.(\ref{small}) ensures
that the contribution to the damping rate of low-energy excitations from the 
$\hat\Psi^{\prime 3}$ terms is much larger than that from the $\hat\Psi%
^{\prime 4}$ terms, and the damping is actually caused by the interaction of
the low-energy excitations with thermal excitations through the condensate
and governed by the processes (\ref{process}). The interaction Hamiltonian
responsible for these processes reads 
\begin{equation}  \label{Hint}
\hat H_{{\rm int}}=\tilde U\int d^3r\Psi_0\hat\Psi^{\prime\dagger} (\hat\Psi%
^{\prime\dagger}+\hat\Psi^{\prime})\hat\Psi^{\prime}.
\end{equation}
Second, under condition (\ref{small}) the damping rate can be found within
the first-order perturbation theory in $H_{{\rm int}}$: 
\begin{equation}  \label{Gamma}
\Gamma_{\nu}\!=\!{\rm Im}\!\sum_{\gamma\gamma^{\prime}} \frac{1}{\hbar} 
\frac{|<\!\!\gamma^{\prime}|\hat H_{{\rm int}}|\nu\gamma\!\!>|^2}{%
E_{\gamma}\!\!-\!E_{\gamma^{\prime}}\!\!+ \!E_{\nu}\!\!+\!i0}
(\!N_{\gamma}\!\!-\!N_{\gamma^{\prime}}\!),\!\!\!
\end{equation}
where $N_{\gamma}=[\exp{(E_{\gamma}/T)}-1]^{-1}$ are equilibrium occupation
numbers for the thermal excitations. The transition matrix element can be
represented in the form 
\begin{equation}  \label{Htr}
\!<\!\!\gamma^{\prime}|\hat H_{{\rm int}}|\nu\gamma\!\!>= \frac{\tilde U}{2}%
\left[ 3H_{\nu\gamma\gamma^{\prime}}\!-\!
(H^{\nu\gamma}_{\gamma^{\prime}}\!-\!H^{\nu\gamma^{\prime}}_{\gamma}
\!\!-\!H^{\gamma\gamma^{\prime}}_{\nu})\right], \!\!
\end{equation}
where $H_{\nu\gamma\gamma^{\prime}}=\int d^3r\Psi_0({\bf r}) f^{-}_{\nu}(%
{\bf r})f^{-}_{\gamma}({\bf r}) f^{-*}_{\gamma^{\prime}}({\bf r})$ and $%
H^{\gamma\gamma^{\prime}}_{\nu}=\int d^3r\Psi_0({\bf r}) f^{-}_{\nu}({\bf r}%
)f^{+}_{\gamma}({\bf r}) f^{+*}_{\gamma^{\prime}}({\bf r})$.

Since energies of the thermal excitations $E_{\!\gamma}\!\!
\gg\!\!\hbar\omega$, these excitations are quasiclassical and, similarly to
the spatially homogeneous case, one can write 
\[
f^{\pm}_{\gamma}({\bf r})=({E_{\gamma}}/ ({\sqrt{E_{\gamma}^2+(n_0({\bf r})%
\tilde U)^2}-n_0({\bf r})\tilde U}) )^{\pm 1/2}f_{\gamma}({\bf r}).
\]
Then, using Eq(\ref{fpm}), from Eq.(\ref{Htr}) we obtain 
\begin{equation}  \label{tr}
\!<\gamma^{\prime}|\hat H_{{\rm int}}|\nu\gamma>=\!\!\left(\!\frac{E_{\nu} 
\tilde U}{2\prod_il_i}\!\right)^{\!\!1/2}\!\!\!\!\!\int\!\!d^3r
\Phi_{\nu\gamma}({\bf r})f_{\gamma}({\bf r})f^{*}_{\gamma^{\prime}}({\bf r}),
\end{equation}
where $\Phi_{\nu\gamma}({\bf r})=W_{\nu}(r_i/l_i) F_{\gamma}({\bf r})$, and 
\begin{equation}  \label{F}
F_{\gamma}({\bf r})\!=\!\frac{2E_{\gamma}^2\!+\!(n_0({\bf r}) \tilde U%
)^2\!-\!n_0({\bf r})\tilde U\!\sqrt{E_{\gamma}^2\!+ \!(n_0({\bf r})\tilde U%
)^2}} {E_{\gamma}\sqrt{E_{\gamma}^2\!+\!(n_0({\bf r})\tilde U)^2}}.
\end{equation}

For the distribution of energy levels of the thermal excitations with a
given set of quantum numbers $\tilde\gamma$ determined by the trap symmetry
(in cylindrically symmetric traps $\tilde\gamma$ is the projection $M$ of
the orbital angular momentum on the symmetry axis) we will use the
statistical Wigner-Dyson \cite{W,D} approach which assumes ergodic behavior
of the excitations. Then, the quantum spectrum of the thermal excitations is
random and the sum in Eq.(\ref{Gamma}) can be replaced by the integral $\int
dE_{\gamma}dE_{\gamma^{\prime}} \sum_{\tilde\gamma\tilde\gamma%
^{\prime}}g_{\gamma} g_{\gamma^{\prime}}R_{\gamma\gamma^{\prime}}$ in which $%
g_{\gamma}(E_{\gamma})$ is the density of states for the excitations with a
given set $\tilde\gamma$, and $R_{\gamma\gamma^{\prime}}$ the level
correlation function. In non-spherical harmonic traps $g_{\gamma}E_{\nu}\gg 1
$ and, hence, $R_{\gamma\gamma^{\prime}}\approx 1$. Then, putting $%
\!N_{\gamma}\!-\!N_{\gamma^{\prime}}=E_{\nu} (dN_{\gamma}/dE_{\gamma})$ and
writing $(E_{\gamma}\!-\!E_{\gamma^{\prime}} \!+\!E_{\nu}\!+\!i0)^{-1}$ as
the integral over time $i\!\int_0^{\infty}\!\!dt\!\exp\{i(E_{\gamma}
\!-\!E_{\gamma^{\prime}}\!+\!E_{\nu}\!+\!i0)t/\hbar\}$, from Eqs.~(\ref
{Gamma}), (\ref{tr}) we obtain: 
\begin{eqnarray}  \label{Gammaint}
\Gamma_{\nu}\!=\!\frac{E_{\nu}^2\tilde U}{2\hbar^2\prod_il_i} {\rm Re}%
\!\!\sum_{\tilde\gamma}\!\!\int\!g_{\gamma} \frac{dN_{\gamma}}{dE_{\gamma}}%
dE_{\gamma} \!\int_0^{\infty}\!\!\!dt\exp{\left\{\!i\frac{(E_{\nu}+i0)t}{%
\hbar} \!\right\}}\!\int\!d^3rd^3r^{\prime}\!\Phi_{\nu\gamma}({\bf r})
\Phi^{*}_{\nu\gamma}({\bf r^{\prime}}) K_{\gamma}({\bf r},{\bf r^{\prime}}%
\!\!,t),
\end{eqnarray}
where the quantummechanical correlation function 
\begin{equation}  \label{K}
\!K_{\gamma}({\bf r},{\bf r^{\prime}}\!,t)\!\!=\!\!\!\sum_{\tilde\gamma%
^{\prime}} \!\!\int\!\!g_{\gamma^{\prime}} dE_{\gamma^{\prime}}\!\exp{%
\!\left\{\!i\frac{(E_{\gamma}\!\!-\!E_{\gamma^{\prime}} )t}{\hbar}\!\right\}}
f_{\gamma}({\bf r})f^{*}_{\gamma}({\bf r^{\prime}})f^{*}_{\gamma^{\prime}}(%
{\bf r}) f_{\gamma^{\prime}}({\bf r^{\prime}}).\!
\end{equation}

In our calculation of the damping rate $\Gamma_{\nu}$ we will turn from the
integration over the quantum states of the quasiclassical thermal
excitations to the integration along the classical trajectories of motion of
Bogolyubov-type quasiparticles in the trap. Following a general method (see 
\cite{Shapoval,Gorkov,Book}), for $K_{\gamma}({\bf r},{\bf r^{\prime}}\!\!,t)
$ we can write 
\begin{equation}  \label{Kq}
\sum_{\tilde\gamma}g_{\gamma}K_{\gamma}({\bf r},{\bf r^{\prime}}\!\!,t)=
\!\!\int\!\! \delta({\bf r^{\prime\prime}}\!\!-\!{\bf r})\delta ({\bf r}_{%
{\rm cl}}(t|{\bf r^{\prime\prime}\!\!,p})\!-\!{\bf r^{\prime}})
\delta(E_{\gamma}\!\!-\!H(p,{\bf r^{\prime\prime}})) \frac{%
d^3pd^3r^{\prime\prime}}{(2\pi\hbar)^3}, \!\!
\end{equation}
where $H(p,{\bf r})\!\!=\!\!\sqrt{(p^2/2m)^2\!+\!2n_0({\bf r}) \tilde U%
(p^2/2m)}$ is the Bogolyubov Hamiltonian, and ${\bf r}_{{\rm cl}}(t|{\bf r,p}%
)$ the coordinate of the classical trajectory with initial momentum ${\bf p}$
and coordinate ${\bf r}$. Then, Eq.(\ref{Gammaint}) is reduced to the form 
\begin{equation}  \label{Gammacal}
\!\Gamma_{\nu}\!\!=\!\!\frac{E_{\nu}^2\tilde U}{2\hbar^2\!\prod_i\!l_i} \!%
{\rm Re}\!\!\int \!\!\!dE_{\gamma}\frac{dN_{\gamma}}{dE_{\gamma}}%
\!\int_0^{\infty}\!\!\! dt\exp{\left(i\frac{E_{\nu}t}{\hbar}\right) }\!\int
\Phi_{\nu\gamma}({\bf r})\Phi^{*}_{\nu\gamma}({\bf r}_{{\rm cl}}(t|{\bf r,p}%
))\delta(E_{\gamma}\!-\!H(p,{\bf r}))\frac{d^3rd^3p} {(2\pi\hbar)^3}.\!\!
\end{equation}

We first consider temperatures $T\gg\mu$, where the main contribution to the
integral in Eq.(\ref{Gammacal}) is provided by the thermal excitations with
energies $E_{\gamma}\alt\mu$. In this case the use of the statistical
approach in non-spherical traps is justified by the fact that, as shown in 
\cite{Graham1}, the motion of corresponding classical Bogolyubov-type
quasiparticles is strongly chaotic at energies of order $\mu$. The
characteristic values of $p$ and $t$ in Eq.(\ref{Gammacal}) are of order $%
(mn_0({\bf r})\tilde U)^{1/2}$ and $\hbar/E_{\nu}$, respectevely. For
quasiclassical low energy excitations ($E_{\nu}\!\gg\!\hbar\omega$) this
time scale is much shorter than $\omega^{-1}$ and important is only a small
part of the classical trajectory, where the condensate density is
practically constant and ${\bf r}_{cl}(t|{\bf r},{\bf p})\!=\!{\bf r}\!+\!%
{\bf v}t$, with ${\bf v}\!=\!\partial H/\partial {\bf p}$. Representing $%
\Phi_{\nu\gamma}({\bf r})\Phi^*_{\nu\gamma}({\bf r}_{cl}(t|{\bf r},{\bf p}))$
as $|f_{\nu}({\bf r})|^2F^2({\bf r})\cos{({\bf p_{\nu}v}t/\hbar)}$, where $%
p_{\nu}$ is a classical momentum of the Bogolyubov-type quasiparticle, from
Eq.(\ref{Gammacal}) we obtain 
\begin{equation}  \label{myGamma}
\Gamma_{\nu}=\int d^3r|f_{\nu}({\bf r})|^2\Gamma_{\nu h}({\bf r}),
\end{equation}
where $\Gamma_{\nu h}({\bf r})$ is the damping rate of the excitation with
energy $E_{\nu}$ in a spatially homogeneous condensate of density $n_0({\bf r%
})$. For $E_{\nu}\ll n_0({\bf r})\tilde U$ we have \cite{Gamma}: $%
\Gamma_{\nu h}=E_{\nu}3\pi^{3/2}(n_{0}a^3) ^{1/2}T/(4n_0\tilde U\hbar)$. A
slight numericall difference from the Szefalusy-Kondor result\cite{SK} is
due to the use of free particle Green functions in their calculation.

For a trapped gas the result of integration in Eq.(\ref{myGamma})
drastically depends on the trapping geometry. For example, for
cyllindrically symmetric harmonic traps $|f_{\nu}({\bf r})|^2\propto(E_{%
\nu}^2+(n_0({\bf r})\tilde U)^2)^{-1/2}$ and strongly increases near the
boundary of the condensate spatial region, where $n_0\tilde U\alt E_{\nu}$.
Just this region of distances determines the density of states $g_M$ and the
damping rate $\Gamma_{\nu}$. Here the contribution to $\Gamma_{\nu}$ from the thermal excitations with $E_{\gamma}$ of order $E_{\nu}$ is important, and one should also include the Beliaev damping processes $\nu\leftrightarrow\gamma+\gamma'$. Omitting the detailes of the calculation which
rely on a generalized version of Eqs.(\ref{tr})-(\ref{Gammacal}) and will be published elsewhere,

 we present here the result for $M=0$: 
\begin{equation}  \label{Gammacyl}
\Gamma_{\nu}=6.7\left(\frac{E_{\nu}}{\mu}\right)^{1/2}\frac{T}{\hbar} \frac{%
(n_{0m}a^3)^{1/2}}{\ln{(2\mu/E_{\nu})}}.
\end{equation}

For the lowest excitations ($E_{\nu}\sim\hbar\omega$) the characteristic
values of $p$ in Eq.(\ref{Gammacal}) are of order $(m\mu)^{1/2}$, and the
result of integration can be represented in the form 
\begin{equation}  \label{Gammafin}
\Gamma_{\nu}=A_{\nu}\frac{E_{\nu}}{\hbar}\frac{T}{\mu}(n_{0m}a^3)^{1/2},
\end{equation}
where $A_{\nu}$ is a numerical coefficient which depends on the form of the
wavefunction of the low-energy excitation $\nu$. In contrast to the case of $%
E_{\nu}\gg \hbar\omega$, the calculation of $A_{\nu}$ requires a full
knowledge of classical trajectories of (stochastic) motion of
Bogolyubov-type quasiparticles in the spatially inhomogeneous Bose-condensed
gas.

The criterion of the collisionless regime for the excitations with energies $%
E_{\gamma}\sim\mu$ assumes that their damping time $\Gamma_{\mu}^{-1}$ is
much larger than the oscillation period in the trap $\omega^{-1}$ and,
hence, the mean free path greatly exceeds the size of the condensate. From
Eq.(\ref{myGamma}) we find $\Gamma_{\!\mu}\!\!
\sim\!\!(T\!/\!\hbar)(\!n_{0m}a^3)^{1/2}$ and obtain the collisionless
criterion 
\begin{equation}  \label{critcol}
(T/\hbar\omega)(n_{0m}a^3)^{1/2}\ll 1.
\end{equation}
Due to collective character of the excitations the criterion (\ref{critcol})
is quite different from the Knudsen criterion in ordinary thermal samples.

Remarkably, both Eq.(\ref{critcol}) and the assumption of stochastic
behavior of thermal excitations with energies of order $\mu$ are well
satisfied in the conditions of the JILA \cite{Jin96} and MIT \cite{Ket97}
experiments, where the temperature dependent damping of the lowest
quadrupole excitations in cylindrically symmetric traps has been measured at
temperatures significantly larger than $\mu$. The JILA experiment \cite
{Jin96}, where the ratio of the axial to radial frequency $%
\beta=\omega_z/\omega_{\rho}=\sqrt{8}$, concerns the damping of two
quadrupole excitations: $M=2$, $E_{\nu}= \sqrt{2}\omega_{\rho}$, and $M=0$, $%
E_{\nu}=1.8\omega_{\rho}$. Our numerical calculation of Eq.(\ref{Gammacal}),
with $W_{\nu}$ from \cite{Ohberg}, gives $A_{\nu}\approx 7$ for $M=2$ and $%
A_{\nu}\approx 5$ for $M=0$. This leads to the damping rate $\Gamma_{\nu}(T)$
which is in agreement with the experimental data \cite{Jin96} (see Fig.1).
In the MIT experiment \cite{Ket97}, where $\beta=0.08$, the damping rate has
been measured for the quadrupole excitation with $M=0$, $E_{\nu}=1.58\omega_z
$. In this case we obtain $A_{\nu}\approx 10$. The corresponding damping
rate $\Gamma_{\nu}(T)$ monotonously increases with $T$ and for the
conditions of the MIT experiment \cite{Ket97} ranges from $4$ s$^{-1}$ at $%
T\approx 200$ nK to $18$ s$^{-1}$ at $T\approx 800$ nK, which is in fair
agreement with the preliminary experimental data.

Importantly, under the condition (\ref{critcol}) the damping rate $%
\Gamma_{\nu}$ of the low-energy excitations is much larger than the damping
rate $\Gamma_T$ of the oscillations of the thermal cloud. This phenomenon
was observed at JILA \cite{Jin96}. One can easily find that for $T\!\gg\!\mu$
the damping rate $\Gamma_T\!\sim\!n\sigma v_T$, where $n\!\sim\!(mT/2\pi%
\hbar^2)^{3/2}$ is the characteristic density of the thermal cloud, $%
\sigma\!=\!8\pi a^2$ the elastic cross section, and $v_T\!\sim\!\sqrt{T/m}$
the thermal velocity. Accordingly, the ratio $\Gamma_T/\Gamma_{\nu}$ is just
of order the lhs of Eq.(\ref{critcol}).

In spherically symmetric traps at any excitation energies one has a complete
separation of variables, which means that the classical motion of
Bogolyubov-type quasiparticles is regular. The excitations are characterized
by the orbital angular momentum $l$ and its projection $M$, and for given $%
l,M$ the level spacing $g_{\gamma}^{-1}\sim\hbar\omega$ can greatly exceed
the interactions provided by the non-Bogolyubov Hamiltonian terms
proportional to $\Psi^{\prime 3}$ and $\Psi^{\prime 4}$. In such a situation
the discrete structure of the energy spectrum of thermal excitations becomes
important, and one can get nonlinear resonances instead of damping. On the
other hand, stochastization of motion of thermal excitations can be provided
by their interaction with each other or with the heat bath. In this case the
damping rate $\Gamma_{\nu}$(\ref{Gammafin}) follows directly from Eq.(\ref
{Gammaint}) by using the Dyson relation for the level correlation function 
\cite{D} ($g_{\gamma}E_{\nu}\!\sim\!1$) and $f_{\gamma}$ from the WKB
analysis of Eq.(\ref{uv}).

For $T\!\!\alt\!\mu$ the picture of damping of low-energy excitations
changes, since $\Gamma_{\nu}$ will be determined by the contribution of
thermal excitations with energies $E_{\gamma}\!\!\!\sim\!\! T$. In this
case, the lower is the ratio $T/\mu$ the more questionable is the assumption
of ergodic behavior of the thermal excitations. But, even if the
stochastization is present, at $T$ significantly lower than $\mu$ the
temperature dependent damping of the lowest excitations will be rather
small. For cylindrically symmetric traps from Eqs.~(\ref{Gammaint}), (\ref
{Gammacal}) one can find $\Gamma_{\nu}\sim (E_{\nu}/\hbar)
(T/\mu)^{3/2}(n_{0m}a^3)^{1/2}$.

This work was supported by the Dutch Foundation FOM, by NWO (project
047-003.036), by INTAS, and by the Russian Foundation for Basic Studies.

\begin{figure}[tbp]
\caption{The damping rate $\Gamma_{\nu}$ versus $T$ for the JILA trapping
geometry. The solid (dashed) curve and boxes (triangles) correspond to our
calculation and the experimental data [6] for the excitations with $M=2$ ($%
M=0$), respectevely.}
\label{1}
\end{figure}


\begin{references}
\bibitem{fn1}  Even in the mean field approach the temperature independent
reversible damping can be obtained \cite{Kagan97,Fantoni}.
\bibitem{f22}  The Szepfalusy-Kondor mechanism was suggested for damping in
trapped gases in the preprints \cite{InternetFolks} which appeared after the
present work has been completed.
\bibitem{Gamma}  The same $\Gamma $ for a homogeneous gas is found in \cite
{InternetFolks}.
\end{references}
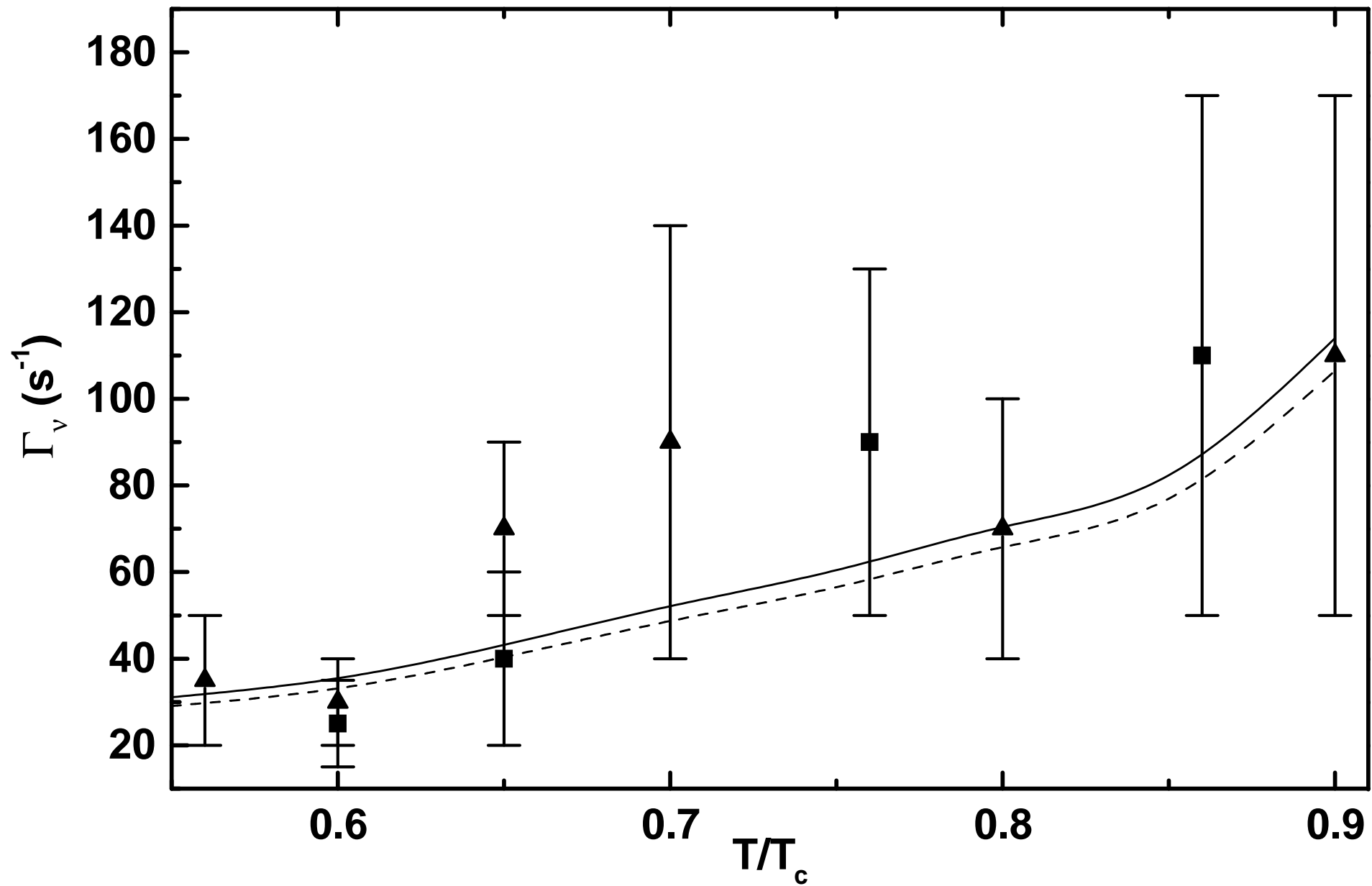
\end{document}